# Inverse designed full-Stokes polarimetric metasurface with simultaneous wavefront sensing for visible light

Ondřej Červinka[1], Martin Hrtoň[1,2], Štěpán Venos[3], Jakub Lelek[3], Libor Úlehla[3], Tomáš Šikola[1,2], Filip Ligmajer[1,2, *]

[1]Brno University of Technology, Central European Institute of Technology, Purkyňova 123, 61200 Brno, Czech Republic

[2]Brno University of Technology, Faculty of Mechanical Engineering, Institute of Physical Engineering, Technická 2, 61669, Brno, Czech Republic

[3]Meopta s.r.o., Kabelíkova 1, 75002 Přerov, Czech Republic

[*]Contact author: filip.ligmajer@vutbr.cz

## Abstract

Metasurfaces have emerged as a powerful platform for compact optical sensors by replacing bulky lenses with flat arrays of subwavelength nanostructures. In precision optical metrology, the simultaneous mapping of a beam's polarization state and wavefront is crucial for real-time diagnostics of stress-induced birefringence and surface flatness. To achieve this in a compact footprint, existing metasurfaces typically partition their aperture into discrete zones, which inherently restricts the light-gathering efficiency and numerical aperture of the system. Here we demonstrate an inverse-designed metasurface that integrates full-Stokes polarimetry and Shack-Hartmann wavefront sensing within a single, continuous aperture in the visible spectrum. By leveraging an adjoint optimization approach to independently control the geometry and rotation of each nanostructure, we break the aperture-sharing paradigm and utilize the entire pixel area for all channels. When coupled with a shallow neural network to automate peak identification and correct for hardware non-idealities, our device yields a mean polarization reconstruction error of only 0.046 across 100 test states on the Poincaré sphere, while simultaneously maintaining the precise focal-spot tracking required for sensitive wavefront tilt detection. This work highlights the capacity of inverse design to generate multifunctional, non-intuitive flat optics that outperforms its traditional counterparts.

## Introduction

Lenses are ubiquitous in many industrial and consumer-electronics applications, and also in the daily life of a scientist, including their use in microscopes, spectrometers, and optical alignment tools. To ensure the high quality of lenses and other optical components, manufacturers subject them to rigorous testing procedures. Among other characteristics, lenses are tested also for stress-induced birefringence and surface flatness. If these properties are not properly controlled, they can negatively impact the optical system's performance by altering the ideal phase profile (wavefront) of the optical component and distorting the polarization state and direction of the transmitted light. Typically, these two quantities are measured in series, by two separate instruments – a polarimeter and a wavefront sensor – increasing costs and leading to longer processing times required for the quality control. Fortunately, modern optics based on nanostructures, embodied by the term metasurfaces, offer novel ways to perform both polarization and wavefront analysis. In polarization analysis, metasurfaces have been already utilized both in a division-of-aperture [1], [2] and division-of-focal plane configuration [3], [4], [5] competing with commercial division-of-time polarimeters in compactness and frame rate of acquisition. The second component of the quality control, wavefront measurements, has also been performed using metasurfaces, replacing microlens arrays with metalenses and providing improved spatial sampling resolution and field of view [6], [7], [8], [9]. In contrast to conventional optics, these two metasurface-based approaches can be merged together for simultaneous acquisition of both the polarization state and wavefront. Such a combination enables real-time observation of internal stresses and deformations of optical components as they are mounted or assembled. Early metasurface implementations of this parallel detection scheme relied on a division-of-aperture approach, using a set of four [10] or six [11] metalenses per pixel to focus individual polarization states. Subsequent designs leveraged the polarization multiplexing technique introduced in [12], which enables a single metalens to focus two orthogonal polarization states simultaneously, effectively doubling the efficiency per zone. However, because analytical design methods struggle to independently map all six polarization bases within a single aperture, these devices still necessitate partitioning the pixel into multiple discrete zones. Consequently, as demonstrated in Refs. [3], [13], the aperture of each metalens remains limited to only one-third of the total pixel size, inherently constraining both the light-gathering efficiency and the numerical aperture of the system. Very recently, polarization-multiplexing has been achieved in the visible domain over all six polarization states in a single zone [14] through the design of meta-molecules consisting of four nanostructures, providing sufficient degrees of freedom to simultaneously address all six polarization basis states. Contrary to the analytical designs used in previous works, the modern inverse design techniques do not require limiting the metasurface layout to zones or metamolecules, but instead they search a vast parameter space of nanostructure layouts to optimize a given cost function [15]. While inverse design approaches have recently shown promise for polarimetric metasurfaces [16], [17] even in the visible range [18], applying them to simultaneously resolve full-Stokes polarization and precise wavefront distortions without sacrificing efficiency or spatial resolution remains an outstanding challenge. Analytical approaches still face performance constraints by dividing the aperture or relying on complex meta-molecules. Consequently, a unified architecture that maximizes light-gathering efficiency while delivering both complete polarization and wavefront metrology without compromises does not exist.

In this work, we present an inverse-designed metasurface that enables simultaneous full-Stokes polarimetry and Shack-Hartmann wavefront sensing in the visible spectrum. As illustrated in Fig. 1a, the device is designed to characterize non-idealities in optical components, such as stress-induced birefringence and surface flatness deviations, which distort both the polarization state and the phase profile of the transmitted light. To address these challenges in a compact footprint, we developed a single-zone metasurface array (Fig. 1b) that transcends the aperture-sharing limitations of traditional zoned designs. By leveraging an adjoint-based inverse design approach, we independently control the size and rotation angle of each nanostructure, enabling the metasurface to focus incident light into six polarization-resolved channels within a single device pixel. As shown in Fig. 1c, the displacement of these focal spots provides a sensitive measure of wavefront tilt, while their intensities are evaluated by a shallow neural network to reconstruct the polarization state of the light. Using a calibration-based mapping from six focal-spot intensities to the Stokes parameters, we demonstrate a mean polarization reconstruction error of 0.05 over 100 test states, while simultaneously measuring wavefront tilt from focal-spot displacement with spatial resolution high enough for precise wavefront metrology.

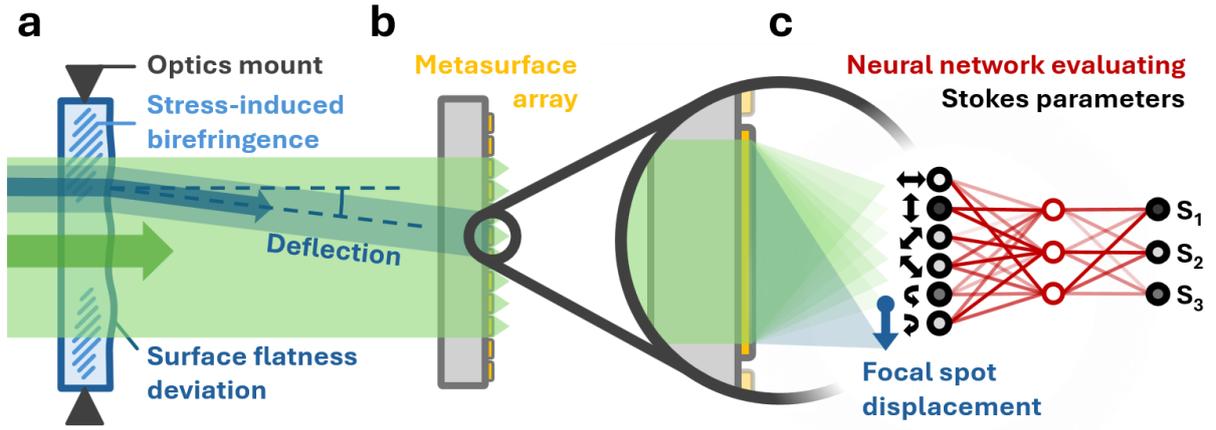

**Fig. 1: Operating concept of our inverse-designed metasurface. a)** Distorted optical component under test, introducing wavefront deflection and stress-induced birefringence. **b)** Metasurface array featuring single-zone polarization multiplexing for simultaneous sensing. **c)** Detail of the detection scheme showing focal spot displacement for wavefront sensing and spot intensities used by a neural network to reconstruct the Stokes parameters.

## Operating Concept and Inverse-Design Strategy

In general, at least four measurements are necessary to fully characterize the polarization state of light through the four components of the Stokes vector [19]. For practical purposes, six measurements are most often made [3], [5], [11], [13], [20], [21], each analyzing one of six elementary polarization states without a decrease in overall efficiency of the final device [1]. The polarization state in a single point of a wavefront can then be fully described by four components of the Stokes vector defined as

$$\vec{S} = \begin{pmatrix} S_0 \\ S_1 \\ S_2 \\ S_3 \end{pmatrix} = \begin{pmatrix} I_H + I_V = I \\ I_H - I_V \\ I_{+45°} - I_{-45°} \\ I_R - I_L \end{pmatrix}, \quad (1)$$

where $I_H$ and $I_V$ represent the intensities of horizontal and vertical linear polarization, $I_{+45°}$ and $I_{-45°}$ the intensities of diagonal linear polarization and $I_R$ and $I_L$ the intensities of right and left circular polarization, respectively.

The wavefront flatness is typically characterized by a Shack-Hartmann sensor [22], which consists of an array of microlenses with focal points in the plane of a photo detector. The larger the angle at which the part of a wavefront hits the respective microlens, the larger the corresponding focal point shift. This effect is preserved when a conventional microlens is replaced by a flat metalens, with the added benefit of decreased shadowing amongst the neighboring lenses, since each metalens is completely flat [23].

To measure the polarization state and the wavefront simultaneously, we designed a metasurface that splits the incident light into six beams according to its polarization state and focuses it into six corresponding focal points. Our metasurface avoids division of the metasurface plane into separate zones by inverse designing the phase profile to cater to all six polarization states at the same time. Compared to zoned architectures that allocate only a fraction of the aperture to each analyzer channel, [3], [4], [5], our single-zone inverse-designed pixel targets all six channels within the full aperture. For our geometry (D = 100 μm, f = 1 mm), this corresponds to NA = 0.05, which is approximately 3× larger than the NA of a 1/3-aperture zoned design at the same focal length. This feature naturally improves their focusing ability at even larger focal distances, which is important for sensitive Shack-Hartmann wavefront sensing. Moreover, it also improves the polarization measurement by improving signal-to-noise ratio per pixel and precision of centroid tracking. Finally, our metasurface also features a square zone design, which enables tiling a plane without empty space in between the device pixels, while sampling the wavefront uniformly in lateral directions. At the same time, the symmetry of pixels translates into the symmetry of focal spots, (compare, e.g., with Ref. [3]), which makes tracking of their positions easier.

Before the optimization began, we pre-computed a library of nanostructures through numerical simulations (see Methods) to get the phase and amplitude response of all nanostructures within the experimentally realizable range. Fig. 2a shows the library we obtained and used for our metasurface optimization, consisting of TiO$_2$ nanostructures with lateral dimensions, width ($W_m$) and length ($L_m$), constrained to 130-330 nm and a height fixed to $H$ = 600 nm. Each entry of the library stores transmission amplitudes ($A$) and phase delays ($\phi$) for $x$- and $y$-polarized light at the target wavelength of $\lambda = 532$ nm.

The initial metasurface layout is generated with random distribution of nanostructure sizes and rotation angles. Each spatially varying nanostructure, indexed by $m$, is then translated into its corresponding Jones matrix elements, $J_{m_x}$ and $J_{m_y}$, by interpolating the values stored in the pre-computed library

$$J_{m_x}(W_m, L_m) = A_{m_x}\exp(i\varphi_{m_x}),$$
$$J_{m_y}(W_m, L_m) = A_{m_y}\exp(i\varphi_{m_y}). \tag{2}$$

To account for the local orientation of the nanostructure at angle $\theta_m$, the corresponding complex transmission matrix $\overleftrightarrow{T}_m$ for the $m$-th nanostructure, is constructed by applying a rotation matrix $\overleftrightarrow{R}(\theta_m)$ to the diagonal Jones matrix as follows:

$$\overleftrightarrow{T}_m = \begin{bmatrix} T_{m_{xx}} & T_{m_{xy}} \\ T_{m_{yx}} & T_{m_{yy}} \end{bmatrix} = \overleftrightarrow{R}(-\theta_m) \begin{bmatrix} J_{m_x}(W_m, L_m) & 0 \\ 0 & J_{m_y}(W_m, L_m) \end{bmatrix} \overleftrightarrow{R}(\theta_m). \quad (3)$$

This parameterization allows the algorithm to treat both dimensions and orientation as continuous variables and allows each element's size and rotation to be independently tuned, maximizing the degrees of freedom available. For each one of the six target input polarization states $\vec{p}_j$, the local transmitted field is obtained simply by multiplying $\vec{p}_j$ with $\overleftrightarrow{T}_m$. In this way, six distinct electric field distributions at the metasurface plane are constructed and then propagated to the far-field focal points using the Green's function formalism [24], yielding the set of six focal point intensities $\{I_{\vec{p}_j}\}$.

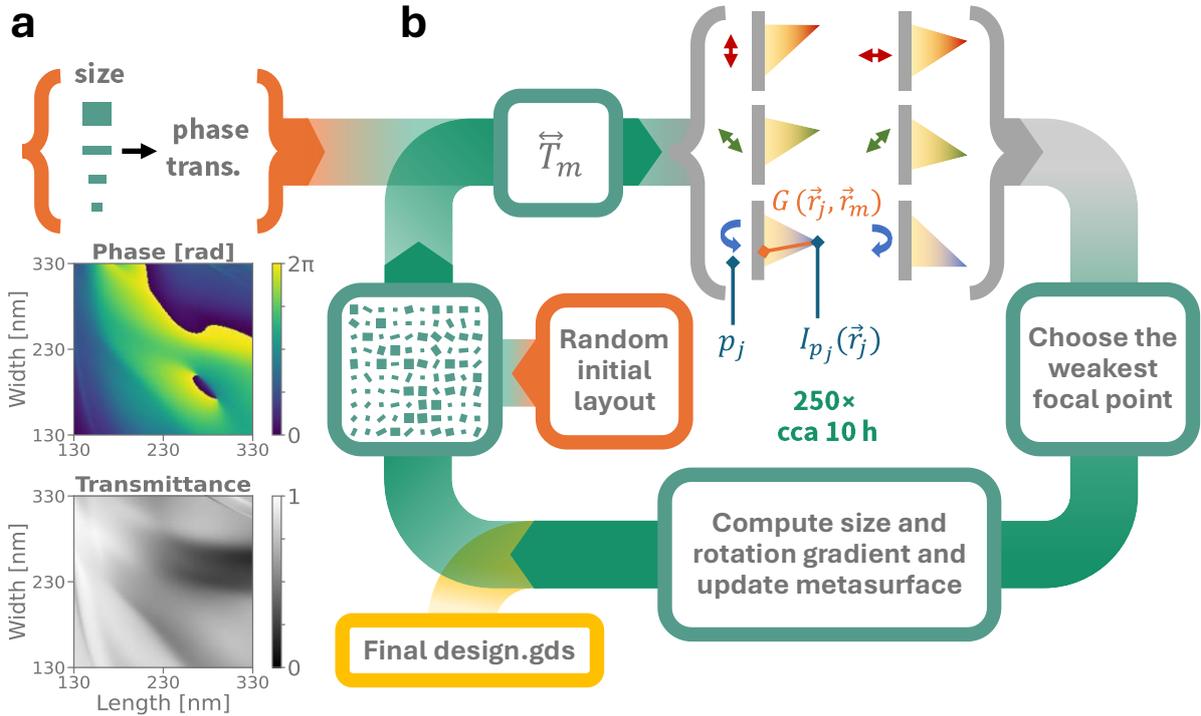

**Fig. 2: Overview of the inverse-design workflow. a)** The pre-computed library of meta-atom building blocks. **b)** Starting from a random design, the algorithm repeatedly translates the metasurface layout into Jones matrices, propagates the near-field into the target focal points, identifies the weakest polarization-resolved focal point at each iteration, computes the corresponding nanostructure size and rotation gradients, and updates the nanostructure layout. After a few hundred iterations, the optimized pattern is exported as a final GDSII layout.

The metasurface is then optimized using an iterative adjoint method algorithm, schematically represented in Fig. 2b, optimizing the cost function

$$\text{maximize } \varepsilon_n = \min_{j \in 1,..,6} \{I_{\vec{p}_j}\}, \quad (4)$$

where $i$ indexes the six target polarization channels, and $j$ identifies the channel with the lowest intensity in the current iteration. At each iteration, the algorithm adjusts the metasurface parameters to increase the intensity $I_{\vec{p}_j}$ at this weakest focal point (targeting polarization state $\vec{p}_j$), gradually balancing and improving the intensities across all six polarization states. The algorithm optimizes the size distribution of nanostructures directly, similar to Refs. [25], [26], while also adding the nanostructure rotation angle as an adjustable optimization parameter. In practice, fabrication constraints may limit the achievable nanostructure geometries, mainly the smallest fabrication-accessible element and the aspect ratio of the nanostructures. Both of these parameters are especially relevant for the future mass fabrication of metasurfaces via nanoimprint lithography. Such constraints may restrict the achievable phase coverage, and consequently the device's final efficiency and completely prohibit forward designs. By working in the real nanostructure size and rotation domain, our inverse design algorithm can realistically accommodate the fabrication-related constraints while maximizing the available phase coverage, even if the full phase-shift range is not accessible.

To know how the nanostructure layout needs to be changed in each iteration, we compute the gradient of the cost function with respect to the geometry of each nanostructure $m$. The geometry parameters $g$ include the nanostructure's width $W_m$, length $L_m$, and rotation angle $\theta_m$. The mathematical expressions for the gradient are summarized in Eqs. (5-8). We begin with the definition of cost function (Eq. (4)) as intensity at the weakest focal point $\vec{r}_n$, written as the scalar product of the electric field components with their complex conjugates at the focal point.

$$\varepsilon_n = |\vec{E}(\vec{r}_n)|^2 = E_x(\vec{r}_n)E_x^*(\vec{r}_n) + E_y(\vec{r}_n)E_y^*(\vec{r}_n). \tag{5}$$

The derivative of the cost function with respect to a geometry parameter $g_k$ then follows as

$$\frac{\partial \varepsilon_n}{\partial g_m} = E_x(\vec{r}_n)\frac{\partial E_x^*(\vec{r}_n)}{\partial g_m} + \frac{\partial E_x(\vec{r}_n)}{\partial g_m}E_x^*(\vec{r}_n) + E_y(\vec{r}_n)\frac{\partial E_y^*(\vec{r}_n)}{\partial g_m} + \frac{\partial E_y(\vec{r}_n)}{\partial g_m}E_y^*(\vec{r}_n), \tag{6}$$

and can be shortened using a mathematical identity for a sum of complex conjugates to

$$\frac{\partial \varepsilon_n}{\partial g_m} = 2\,Re\left\{E_x^*(\vec{r}_n)\frac{\partial E_x(\vec{r}_n)}{\partial g_m} + E_y^*(\vec{r}_n)\frac{\partial E_y(\vec{r}_n)}{\partial g_m}\right\} = 2\,Re\left\{\vec{E}^*(\vec{r}_n) \cdot \frac{\partial \vec{E}(\vec{r}_n)}{\partial g_m}\right\}. \tag{7}$$

The change in the far-field electric field at the focal point can be related to changes in the metasurface near-field evaluated closely above the metasurface plane, by propagation to the focal point via the Green's transfer function $G$ from the metasurface building block at $\vec{r}_m$ to a focal point at $\vec{r}_n$:

$$\frac{\partial \varepsilon_n}{\partial g_m} = 2\,Re\left\{\vec{E}^*(\vec{r}_n) \cdot \sum_{\vec{r}_m} G(\vec{r}_n, \vec{r}_m)\frac{\partial \vec{T}_m}{\partial g_m}\vec{p}_n\right\}, \tag{8}$$

where $\vec{p}_n$ represents polarization state targeted by the worst performing lens. Examples of this derivative expressed for specific geometry parameters are provided in the Supporting Information.

After the optimization converges to a stable solution, where all the focal points have the same intensity and are no longer improving, we export the final nanostructures' size and rotation distribution into a GDSII file for subsequent nanofabrication. The convergence usually occurs after a few hundred iterations and can be assisted by gradually decreasing the gradient step – the maximum change applied to the nanostructure size and rotation in each iteration. At every iteration, the raw (adjoint-computed) gradients of size and rotation are normalized so that their largest absolute value matches the chosen gradient-step value. Thus, even if the gradient magnitude varies strongly during the optimization, the applied update is bounded by the selected step size. To ensure stable convergence, this gradient step is reduced linearly throughout the optimization. Specifically, in our implementation, the size and rotation steps decrease from 10 nm/iteration and 0.3 rad/iteration, respectively, to zero in the final iteration. Middle section of the optimized metasurface is shown in Fig. 3a, showing the counter-intuitive nature of the inverse optimized design. Evaluations of the simulated electric field distributions for this final optimized layout demonstrate a near-perfect polarization reconstruction capability, achieving an average error on the order of $10^{-7}$ across the Poincaré sphere.

We fabricated the optimized metasurface by a combination of electron beam lithography and atomic layer deposition, forming high-aspect-ratio $TiO_2$ nanostructures on a silica substrate, following the approach pioneered by the Capasso's group [27] (see Methods). Total area of one metasurface pixel is 100×100 μm, and due to its square shape, it can uniformly tile the whole Shack-Hartmann sensor surface. The SEM micrograph of the fabricated device is shown in Fig. 3b.

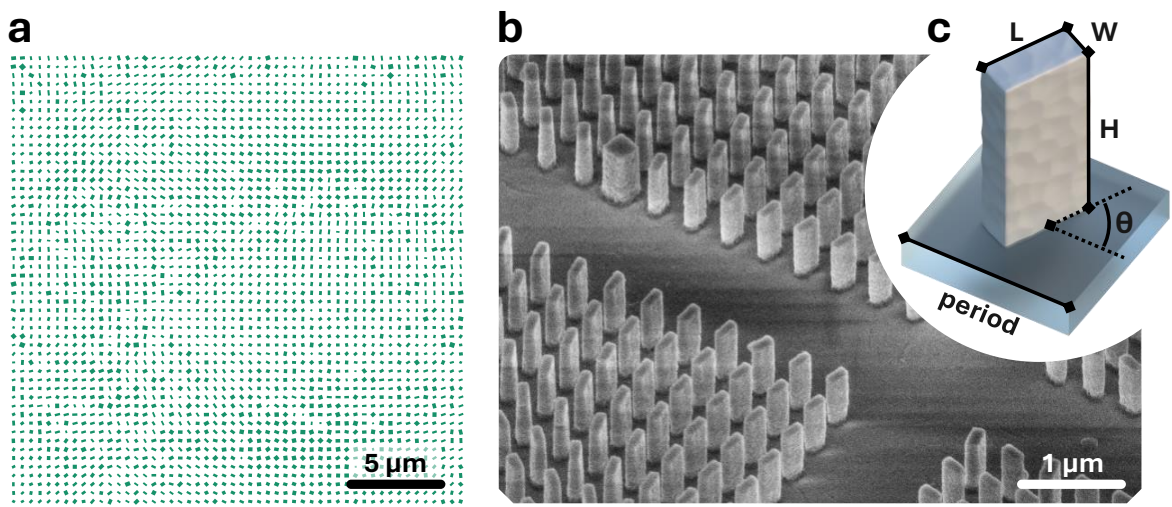

**Fig. 3: Fabricated metasurface nanostructures. a)** Top-view schematic of the metasurface center showing the counter-intuitive design. **b)** Micrograph at a corner of the periodically tiled metasurface. **c)** Individual nanostructure dimension designation.

## Results and discussion

The functionality of the fabricated device in terms of polarization characterization was measured using a custom-built optical setup, schematically represented in Fig. 4a. This setup allows for testing the metasurface's response to various incident polarization states prepared by automatically rotating a half-wave plate and a quarter-wave plate in combination with a stationary polarizer. Using this system, we measured the device's response to one hundred polarization states uniformly distributed over the Poincaré sphere. The resulting focal spots were recorded with a camera, and the intensity of each focal point was estimated by summing the pixel values within a fixed radius around the peak centers, as illustrated in Fig. 4b. Standard polarimetric reconstruction using Eq. (1) assumes ideal, orthogonal filters. However, in any high-NA multiplexed system, non-idealities such as crosstalk and unequal power distribution across non-orthogonal channels can introduce errors [28]. To account for these system-level complexities, we reframe the polarization evaluation as an inference task for a shallow neural network that maps the six measured focal spot intensities on the input to the four Stokes parameters on the output. The use of shallow neural network is justified by its ability to correct for subtle experimental non-idealities, such as CMOS sensor non-linearity, which a standard linear model may fail to capture [29], [30]. Furthermore, the network architecture inherently automates the identification of focal peaks and the subtraction of non-uniform backgrounds. A shallow design provides sufficient flexibility to handle these hardware-specific artifacts while preventing overfitting, ensuring robust and generalizable reconstruction. By training the network on the measured data collected with the optical setup, we effectively calibrate the system against experimental non-idealities including fabrication imperfections. The resulting shallow neural-network predictions for the one hundred test images, compared with the known incident polarization states set by the waveplates, are shown in Fig. 4c. We extracted the average inference error (defined as the Euclidean distance between the true and predicted Stokes parameters) across the whole Poincaré sphere to be 0.046. Using the standard linear calibration matrix approach yields slightly higher average error of 0.049. Note that we do not normalize the final predictions and thus allow ourselves to also infer polarization states of partially polarized light. The error rate of our inverse designed polarimeter is in-line or better than the values reported for imaging polarimeters based on division-of-image approach (0.122 @ 550 nm, Ref. [18]), plasmonic metagratings (0.112 @ 800 nm, Ref. [31]) and dielectric spatial multiplexing metasurfaces (0.075-0.130 @ 800 nm, Ref. [3]) or on previous inverse design approach (0.059 @ 10.6 µm, Ref. [16]). To the best of our knowledge, the only superior approach reported until now has been a disordered-metasurface approach (0.009 @ 550 nm, Ref. [32]), which, however, does not conceptually allow for wavefront sensing. The overall power efficiency of our inverse-designed metasurface is the same as that of an equivalent metasurface designed with three separate zones using spatial multiplexing, which has been verified by integrating the power over the focal points in simulations of the two designs. However, the focal spots produced by the inverse-designed metasurface were significantly better confined, allowing the integration to be performed over a smaller area and improve the dynamic range of Shack-Hartmann wavefront sensing.

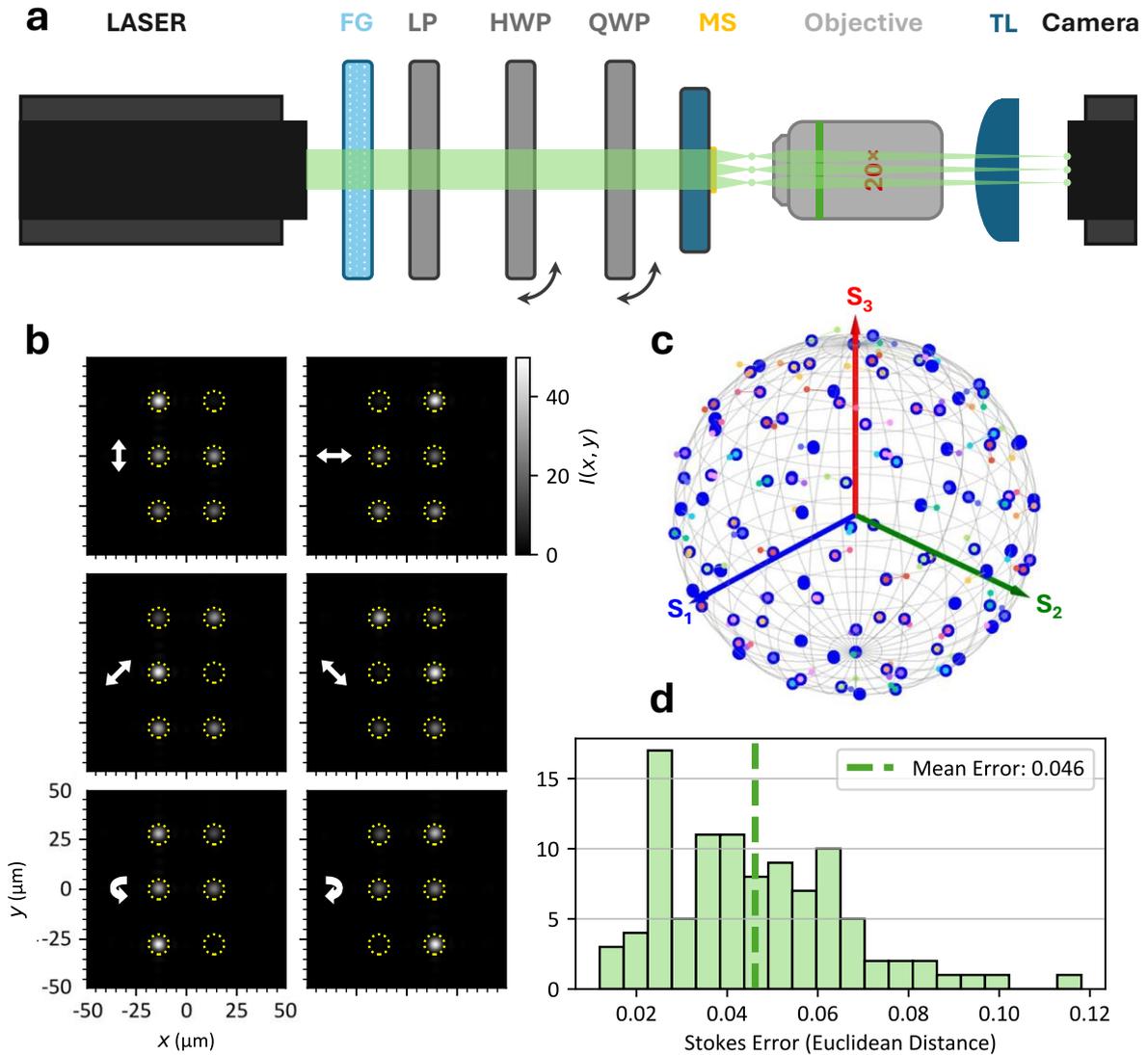

**Fig. 4: Polarization characterization. a)** Polarization measurement setup and results (FG-frosted glass, LP-linear polarizer, HWP-half-wave plate, QWP-quarter-wave plate, MS-metasurface, TL-tube lens). **b)** Simulated intensity spots distribution for the six polarization bases states (labeled by the white arrows). **c)** Poincaré sphere showing the target polarization states (blue spots) connected with the polarization state estimated by the neural network. **d)** Histogram of Stokes vector estimation error magnitudes.

To characterize the Shack–Hartmann wavefront-sensing performance of our metasurface, we built a custom optical setup shown in Fig. 5a. The setup includes a rotatable arm carrying a laser source and a rotating frosted glass diffuser, which reduces temporal coherence and suppresses speckles in the image plane. The metasurface was placed directly in the center of rotation of the stage which allowed us to achieve wide range of incident illumination angles without the need for sample repositioning. The camera (Moravian Instruments C1+12000A with 3.45 μm pixel size) then captured the positions of the focal points within the image plane as a function of the incident angle set by the rotatable stage. The measured data, plotted in Fig. 5b, show a displacement of 126.1 px/deg on the sensor. Given the sensor pixel pitch of

3.45 μm, this corresponds to a measured sensitivity of approximately 435 μm/deg at the camera plane. Accounting for the 20× magnification of the imaging system, this translates to a physical sensitivity of approximately 21.8 μm/deg within the metasurface focal plane. In a Shack-Hartmann sensor, the sensitivity of wavefront measurements is a fundamental trade-off against the allowable range of incident angles, constrained by the numerical aperture of the individual lens, which must be chosen to avoid overlaps between neighboring focal spots. In our metasurface-based implementation, the spot size and spacing are similarly critical: too large a spot or too small a separation could lead to overlaps, reducing measurement feasibility. Our design produces focal spots with a FWHM of around 4 μm, close to the diffraction-limited 2.7 μm, and spaced 25 μm apart to maintain clear separation. Having multiple focal points, which is necessary for polarization detection, however, decreases the effective angle range for wavefront measurement, before the focal spots overlap with those of neighboring zones, to 1.4° (corresponding to the focal distance of 1 mm and separation between focal spot and edge of 25 μm). This range is sufficient for measurement of typical planar distortions and can be increased further by packing the focal spots closer, which is permitted by the increased NA of the inverse-designed lens, compared to spatial multiplexing designs.

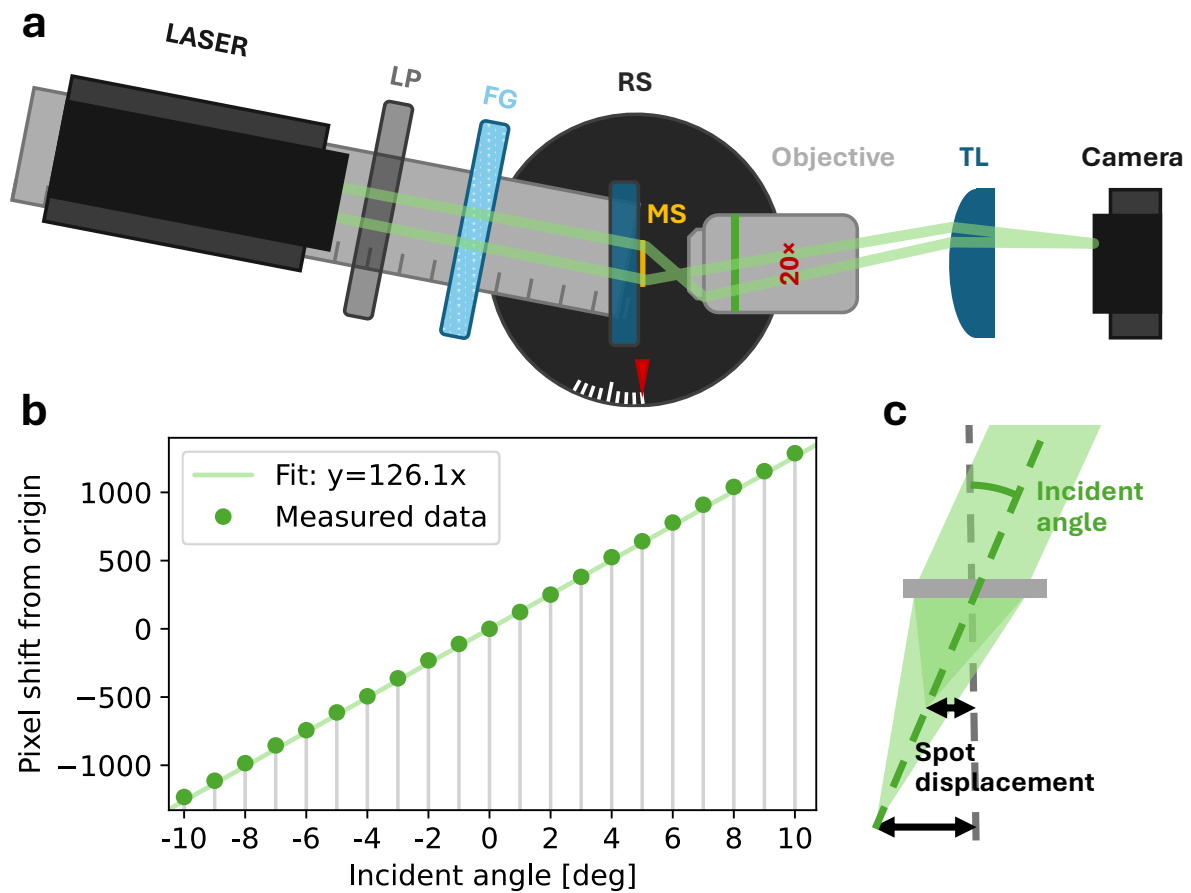

**Fig. 5: Shack-Hartmann measurement. a)** Optical setup for wavefront measurement (LP-linear polarizer, FG-frosted glass, RS-rotating stage, MS-metasurface, TL-tube lens). **b)** Measured focal spot displacement as a function of incident angle with linear interpolation. **c)** Illustration of the principle of NA vs. sensitivity trade-off.

## Conclusion

We have demonstrated an inverse-designed, visible-light metasurface that integrates full-Stokes polarimetry and Shack-Hartmann wavefront sensing into a single, compact aperture. By moving beyond the division-of-aperture paradigm, our single-zone design utilizes the entire pixel area for every polarization channel, significantly increasing the device's numerical aperture. The resulting device achieves high-fidelity polarization reconstruction, yielding a mean error magnitude of just 0.046 over 100 test states, while maintaining the sensitive wavefront tilt detection required for high-resolution optical metrology. Such a combination of functionalities within a compact, tile-able footprint showcases the capacity of inverse design to create highly integrated optical sensors for real-time diagnostics of stress and surface-flatness in optical elements. This work shows how optimization-driven meta-optics can outperform traditional analytical designs in both multi-functionality and precision.

## Methods

Simulations:

We simulated the optical response of the nanostructures using rigorous coupled-wave analysis (RCWA) implemented in the Python Inkstone library [33], employing the method's inherent periodic boundary conditions with a period of $p = 450$ nm. The lateral dimensions of the nanostructures varied from 130 to 330 nm. The nanostructures were modeled as $TiO_2$ on a $SiO_2$ substrate. The refractive index of $TiO_2$ was obtained from ellipsometric measurements of an ALD-deposited film, while the fused silica substrate refractive index was set to 1.44. The simulations were performed at a target wavelength of 532 nm, with the number of spatial frequency components being set to 300 after convergence testing.

Fabrication:

The high aspect ratio $TiO_2$ nanostructures were fabricated using a method combining electron beam lithography with atomic layer deposition. First, a fused silica wafer (10×10 mm chip) is spin-coated with a resist. Because the thickness of the resist determines the final height of the nanostructures, it is necessary to select a viscous enough resist. The used CSAR 6200.13 results in a 650 nm layer if spin-coated at 1600 RPM. An additional thin layer of chromium (5 nm) was deposited on top to serve as a charge dissipation layer. The nanostructure layout is then transferred into the resist via electron beam lithography. We utilized a 30 kV e-beam system RAITH 150 Two for the exposure, with a dose of 170 µC/cm². The exposed resist was then stripped of the chromium layer and developed using a standard developer amyl acetate for a duration of one minute. The developed resist mask was then transferred into an atomic layer deposition chamber, where it was conformally filled with a 170 nm $TiO_2$ layer. The ALD recipe was optimized for 90 °C not to melt the e-beam resist mask. 2600 cycles of alternating tetrakis(dimethylamino)titanium and water precursors deposit the 170 nm layer in about 13 hours. The ALD deposition, however, not only fills the resist mask hollows but also creates a layer over the whole sample surface that needs to be removed. To remove the top layer, we etched the sample surface with an argon ion beam in Scia Systems Coat 200. This system is also equipped with a secondary ion mass spectrometer, which allows us to stop the etching precisely when the top layer is just removed. In the final step, the sample is put into a Diener resist stripper, where the resist mask is removed in oxygen plasma within 150 minutes. This leaves the high aspect ratio $TiO_2$ nanostructures free-standing on the substrate.

Neural-network Stokes vector inference:

Stokes parameters were reconstructed using a shallow neural network (6-3-3) implemented in PyTorch. The network employed a ReLU-activated hidden layer to map six input intensities to the $S_1$, $S_2$ and $S_3$ components. The model was trained using the Adam optimizer to minimize mean squared error over 10 000 iterations (≈5 s on a modern CPU), and its performance was benchmarked against a traditional linear least-squares calibration.

# Acknowledgment

This work was supported by the Grant Agency of the Czech Republic (project No. 25-18336M). We acknowledge CzechNanoLab Research Infrastructure (ID 90251), funded by MEYS CR, for the financial support of the sample fabrication.